# ActiveAI: Enabling K-12 AI Literacy Education & Analytics at Scale

Ruiwei Xiao[1], Ying-Jui Tseng[1], Hanqi Li[2], Hsuan Nieu[3], Guanze Liao[3], John Stamper[1], Kenneth R. Koedinger[1]

[1]Carnegie Mellon University  [2]New York University  [3]National Tsing Hua University

ruiweix@andrew.cmu.edu

**ABSTRACT**: Interest in K-12 AI Literacy education has surged in the past year, yet large-scale learning data remains scarce despite considerable efforts in developing learning materials and running summer programs. To make larger scale dataset available and enable more replicable findings, we developed an intelligent online learning platform featuring AI Literacy modules and assessments, engaging 1,000 users from 12 secondary schools. Preliminary analysis of the data reveals patterns in prior knowledge levels of AI Literacy, gender differences in assessment scores, and the effectiveness of instructional activities. With open access to this de-identified dataset, researchers can perform secondary analyses, advancing the understanding in this emerging field of AI Literacy education.

## 1   INTRODUCTION

K-12 AI literacy education has gained significant attention in the past year (Klopfer et al., 2024). While researchers have made considerable progress in designing learning materials and organizing summer camps, large-scale learning platforms (Tseng et al., 2024) and datasets (Almatrafi et al., 2024) remain limited. To provide accessible AI literacy learning materials for schools, as well as scalable datasets to support replicable research in the learning analytics community, we developed a K-12 AI Literacy learning platform. This platform offers evidence-based learning activities and assessments for classroom use, along with standardized data logging compatible with widely-used educational data repositories for secondary analysis. Over the past year, our efforts in instructional design, platform development, and school partnerships have resulted in the collection of AI literacy learning and assessment data from over 1,000 users across 4 learning modules.

## 2   METHODS

### 2.1   System Design and Data Pipeline

The platform is implemented as a web application developed using Next.js and OpenAI APIs, supporting three types of users (**students, teachers, researchers**). Example interfaces and data flows for each role are illustrated in Figure 1. In a complete learning experience, **students** generate all the data, and they have access only to their own data. Students begin by completing a survey to provide de-identified demographic information, followed by a sequential process of a pre-test, learning module, and post-test on the assigned topic. The pre- and post-tests are isomorphic assessments targeting the same learning objectives, while the learning module includes interactive activities with an AI agent in simulated real-life scenarios, such as identifying LLM hallucinations in news summaries, to teach AI literacy learning objectives. Learners' interaction data from learning activities and assessments are logged into the learning management system and aggregated at varying levels of granularity for other user roles. **Teachers** can access survey and activity data from their classes. Specifically, teachers can view data on time spent, completion rates, and correctness, aggregated by activity, learning objective, module, student, and class. This real-time data aggregation allows teachers to make informed adjustments to their instructional plans and provide timely support to students in need. **Researchers** can access all types of de-identified data and their aggregated forms. Instead of providing interfaces and visualizations, the platform supports downloading standardized data logs compatible with widely-used educational data repositories (e.g., DataShop[1], LearnSphere[2]) and data analysis tools (e.g., RStudio, Tableau) for public access and analysis. With different levels of aggregation, our data supports learning analytics, including learner modeling, validation of existing learning sciences principles, and learning engineering within AI literacy as a new domain.

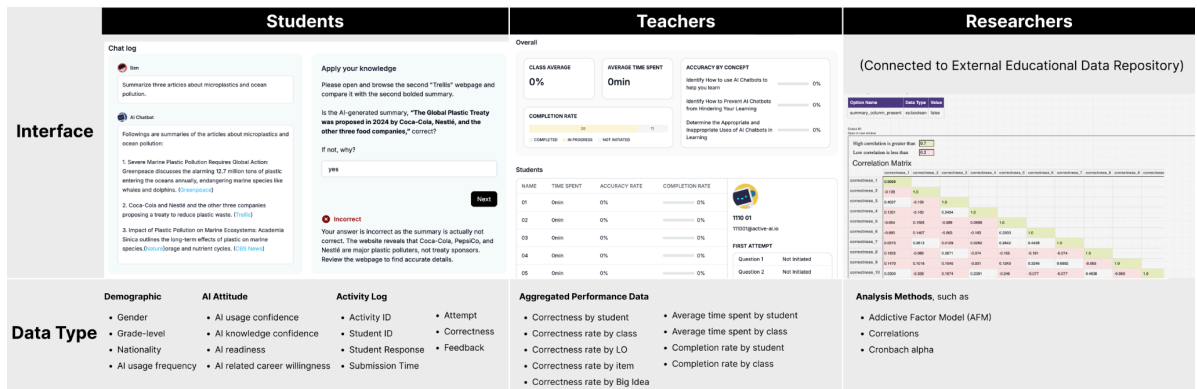

Figure 1: System Design and Data Pipeline

## 2.2 Learning Design on K-12 AI Literacy Modules

We use a backward design approach (Wiggins & McTighe, 1998) to create learning objectives, assessments, and activities aligned with AI literacy standards (Touretzky et al., 2023) and priority topics from partner schools. For example, to address teachers' interest in identifying LLM hallucinations, we map this skill to AI4K12's Big Idea #5: Societal Impact, design corresponding assessments, and develop interactive, scaffolded activities with feedback (Figure 1). Empirical examples are detailed in our prior work (Tseng et al., 2024).

## 3 PRELIMINARY RESULTS AND FUTURE WORKS

We partnered with 12 secondary schools across North America, Asia, and Australia. Eight learning modules were implemented, with four deployed in four schools along with surveys and pre- and post-tests. Over 1,000 unique learners used the platform, and 426 (171 for modules 1 and 2, 131 for module 3, and 114 for module 4) completed all components, providing complete data for analysis. To triangulate the results and enhance the interpretability of our findings, we conducted teacher interviews and student cognitive task analysis. Preliminary results indicate learning gains, gender differences, and variations across educational contexts.

---

[1] DataShop: https://pslcdatashop.web.cmu.edu/

[2] LearnSphere: https://learnsphere.org/

Across the 7 learning objectives in 4 modules, we observed significant learning gains in 4 of them, based on Wilcoxon tests on pre- and post-test scores on the non-normal distributed data. Learning gains correlated with cognitive engagement levels (ICAP framework, Chi & Wylie, 2014): objectives with significant gains involved interactive activities, while those with smaller gains were linked to passive reading. This highlights the importance of cognitive engagement, though further analysis is needed to identify areas for improvement.

In the latest experiment involving module 4 (Identify LLM Hallucinations), where gender data is available, non-male students achieved significantly higher assessment scores on both the pre-test **(f=6.97, p<0.01, one-way ANOVA)** and post-test **(f=6.80, p=0.01, one-way ANOVA)** compared to their male counterparts, along with slightly higher learning gains. These findings help researchers identify threats to activities validity and improve them through targeted interventions. In the future, as 2 partner schools implement 4 learning modules by year-end, we will collect AI literacy data over a longer duration. Additionally, while Asian schools offer standalone IT courses dedicated to AI literacy, schools in other regions integrate our materials into regular subjects or STEM clubs. These differences in learning contexts will also enrich our dataset.

## 4 CONTRIBUTIONS

This study makes three key contributions to the LAK community: 1. **A Novel Dataset**: We provide a dataset in the emerging and understudied domain of AI literacy, capturing learners' prior knowledge, interactions with LLM systems, and learning outcomes, enriched with demographic and contextual information. 2. **Open Access for Secondary Analysis**: The dataset will be made openly available on established educational data repositories, offering a valuable resource for secondary analysis and enabling broader research fields. 3. **Practical AI Literacy Resources:** A set of AI literacy learning activities are provided to support practitioners in integrating AI concepts into diverse classroom environments. These contributions aim to advance knowledge, research and practice in AI literacy education, fostering a deeper understanding and scalable approaches through learning analytics.